\documentstyle[12pt]{article}

\def\sds{D\!\!\!\!\not\,\,\,\,\,}

\def\sds{D\!\!\!\!\not\,\,\,\,\,}
\def\endpage{\vfill\eject}
\begin{document}
\vskip 2. truecm
\centerline{\bf
The Schwinger Model on the lattice in}
\centerline{\bf
the Microcanonical Fermionic Average  approach.}
\vskip 2 truecm
\centerline { V.~Azcoiti$^a$, G. Di Carlo$^b$, A. Galante$^{b,c}$,
A.F. Grillo$^b$ and V. Laliena$^a$}
\vskip 1 truecm
\centerline {\it $^a$ Departamento de F\'\i sica Te\'orica, Facultad de
Ciencias, Universidad de Zaragoza,}
\centerline {\it 50009 Zaragoza (Spain).}
\vskip 0.15 truecm
\centerline {\it $^b$ Istituto Nazionale di Fisica Nucleare, Laboratori
Nazionali di Frascati,}
\centerline {\it P.O.B. 13 - Frascati (Italy). }
\vskip 0.15 truecm
\centerline {\it $^c$ Dipartimento di Fisica dell'Universit\'a dell'Aquila,
67100 L'Aquila, (Italy)}
\vskip 2 truecm

\begin{abstract}
The Microcanonical Fermionic Average method has been used so far in the
context of
lattice models with phase transitions at finite coupling. To test its
applicability to Asymptotically Free theories, we have implemented it in
QED$_2$, \it i.e.
\rm the
Schwinger Model.
We exploit the possibility, intrinsic to this method,
of studying
the whole $\beta, m$ plane at negligible computer cost,
to follow constant physics trajectories and
measure the $m \to 0$ limit of the chiral condensate.
We recover
the continuum
result within 3 decimal places.
\end{abstract}
\endpage
The Microcanonical Fermionic Average
(M.F.A.) method for performing Lattice
simulations with dynamical fermions
\cite{GEN} is ideally suited for discussing the
phase structure of theories with phase transitions at finite couplings, and
it has been applied so far in this context \cite{QED4,QED3}.

The conventional wisdom, however, requires that physically interesting
theories are Asymptotically Free like QCD. It is
then interesting to test the applicability of the M.F.A. method to a theory
without phase transitions at finite coupling \cite{QCD3}.
In this paper we present an analysis of the Schwinger Model on the lattice.
Strictly speaking, the Schwinger Model in the continuum is not Asymptotically
Free, since it is superrenormalizable and
the Callan-Symanzik $\beta$ function vanishes.
However, in the lattice version, since the continuum coupling is
dimensionful, the continuum theory is reached at infinite lattice coupling,
much in the same way as four dimensional Asymptotically Free theories like
QCD.

The continuum model is confining; it is
exactly solvable at zero fermionic mass, so
that we can compare the results of our simulations with
exact ones.
We have simulated the (unquenched)
model in lattices ranging from $16^2$ to $100^2$; we present here results
for the average plaquette and for the chiral condensate,
in the non symmetric ($\theta=
0 $) vacuum of the model.

The evaluation of the chiral condensate has been made easier by the fact
that, in the M.F.A. approach, the main computer
cost resides in the evaluation of an effective fermionic action at fixed
pure gauge energy by evaluating {\it all} the eigenvalues of the fermionic
matrix at $m=0$. It is then essentially possible, at no extra
cost, to move in the plane $\beta, m$ to follow constant physics trajectories
in approaching the correct continuum limit. This is easier in this model
since here the Renormalization Group amounts to simple dimensional analysis.

The M.F.A. method is fully described in \cite{GEN}.
Starting from the partition function
in terms of the total Action $S = S_F+S_G,$
sum of the fermionic and
pure gauge contributions, we define the density of states at fixed
pure gauge Action ({\it i.e.} Euclidean Energy)
\begin{equation}
N(E)= \int DU \delta(S_G(U)-VE)
\end{equation}
and an effective fermionic action through
\begin{equation}
e^{-S_{eff}^F(m,n_f,E)}=\langle \det\Delta^{n_f \over 2}\rangle _E =
{{\int DU \det\Delta^{n_f \over 2}\delta(S_G(U)-VE)} \over {N(E)}}
\end{equation}
which is the microcanonical average
of the fermionic determinant.

In terms of the effective Action the partition function can thus be
rewritten as
\begin{equation}
{\cal Z} = \int dE N(E)e^{- \beta VE-S_{eff}^F(m,n_f,E)}
\end{equation}

Massless electrodynamics in $1+1$ dimensions
is confining,
superrenormalizable and exactly solvable.

Its partition function is
\begin{equation}
{\cal Z}=
\int DA_\mu D\bar\psi D\psi e^{\int d^2x[{1 \over 4} F_{\mu\nu} F_{\mu\nu}
+\bar\psi \sds\psi]}
\end{equation}
with the usual definitions of $F_{\mu\nu}$ and $\sds$. The electric charge is

The partition function (in the photonic sector)
can be rewritten as \cite{REUTER}
\begin{equation}
{\cal Z} =\int DA_\mu e^{\int d^2x[{1 \over 4}
F_{\mu\nu}F_{\mu\nu}+{e^2 \over 2\pi} A_\mu A_\mu]}
\end{equation}
{\it i.e.} as that of a theory of free massive vector bosons
of mass $M = {e \over \sqrt {\pi}} $.
In particular the Green's functions of purely
bosonic operators are the same in both theories.
This fact can be exploited for obtaining the
average plaquette in the lattice (see later).

As for the chiral properties of the model, the chiral current is anomalous.
If the chiral limit is obtained from
$m \ne 0$, then the $\theta=0$ vacuum is selected.
In this vacuum the chiral condensate is (with one flavour)

\begin{equation}
{1 \over e}\langle \bar\psi\psi\rangle_c = {e^{\gamma_e}\over {2 \pi
\sqrt \pi}} = 0.15995
\end{equation}
 while it diverges at zero flavour ({\it i.e.} the quenched limit) and is
zero with two flavours.
This is the value of the chiral condensate to be compared with the results
of lattice simulations, where its chiral limit is obtained from $m \ne 0$

In the present simulation the pure gauge part is described in terms of
{\it non compact} fields, while
for the fermionic - gauge term we use the standard
staggered formulation with $n_f$ species.

Since the photonic sector of the continuum theory is equivalent to a theory
of a free, massive vector
boson, the
average plaquette of the Schwinger model can be compared with
that of the vector boson, which can
be exactly computed on a finite lattice:
\begin{equation}
\langle E \rangle_L={ 1 \over 2V}
\sum_{p_1 p_4} { 2-\cos p_1-\cos p_4 \over 2\beta
\sum_\gamma (1-\cos p_\gamma)+ M^2}
\end{equation}
$(p_\mu={2\pi \over Na} k_\mu) $
and for $V \to \infty$
\begin{equation}
E={1 \over 2}\int {d^2p \over (2\pi)^2}{ 2(1-\cos p_1)+2(1-\cos p_4)
 \over M^2+2\beta \sum_\gamma (1-\cos p_\gamma)}
\end{equation}
The value $M = { 1 \over \sqrt\pi}$ corresponds to the continuum
Schwinger model,
while the quenched value is
\begin{equation}
E(M=0)= { 1 \over 2\beta}
\end{equation}
 Since
$e_c$ is dimensionful, $\beta$ explicitely contains the lattice spacing:
$\beta={1 \over a^2 e^2_c}$ so that the
continuum limit of the theory is approached at $\beta \to \infty$.
The limit must be reached keeping fixed the
dimensionless ratio ${m_c \over e_c} = {\sqrt \beta} m$.
This ratio defines constant physics trajectories.

We have performed simulations
in lattices up to $100^2$. We present here the
results for the $64^2$ lattice, where we have the best statistics
(for a total of $70$ Cray-equivalent hours) \cite{SCHL}. We will
mainly discuss the $1-$flavour case.

As stated before, we compute all the eigenvalues of the fermionic matrix.
This allows us to compute the Effective Action for all values of the mass,
including $m = 0$. We have done so for $20$ values of the energy, from
$0.08$ to $1.3$.

One advantage of the MFA method is that the phase structure of the theory
can be inspected directly from the fermionic effective action, whose
derivatives must be discontinuous in order to generate a phase transition,
at least for small $n_f$,
if the underlying pure gauge theory has no transition \cite{QED4}.
In the case of QED$_2$ the continuum theory is obtained as $\beta \to
\infty$ and one does not expects finite $\beta$ transitions. The effective
fermionic action numerically evaluated for the model does not show any sign
of non analyticity and hence of phase transition \cite{SCHL}.

The average plaquette is obtained as
\begin{equation}
\langle E \rangle_L =
{\int dE N(E) E e^{- \beta VE}e^{-S_{eff}^F(m,n_f,E)}\over
{\cal Z}}
\end{equation}
and can be directly computed at $m=0$. Since the underlying pure gauge
theory is quadratic the density of states is known analytically
\begin{equation}
N(E)=C_GE^{{1 \over 2}V-{3 \over 2}}
\end{equation}
so that the integrals in (10) are simple one-dimensional integrals.

In Figure 1 we report the value of the average plaquette
energy (diamonds), multiplied by $2 \beta$ to improve the visibility, compared
with the exact result for a Massive Vector Model on the lattice. It is
important to notice that the Schwinger Model is equivalent to a Vector Model
in the continuum. On the lattice, there is no guarantee that the two models
are related. From Figure 1 one can see that at small $\beta$, where
presumably
we are far from the continuum, there is disagreement between the numerical
results and the analytical ones. However, already at $\beta \sim 1$ the
agreement becomes excellent, showing that, at least for this operator, the
continuum physics is reached fastly.
The straight line is the quenched value $2\beta \langle E \rangle = 1$, and
one can see that, as $m$ increases, the asymptotic value of $\langle E
\rangle$ moves towards it.

The chiral condensate
\begin{equation}
\langle\bar\psi\psi\rangle=-{1 \over n_f V} {\int dE e^{-S_{eff}} {\partial
\over \partial m} S^F_{eff} \over \int dE e^{-S_{eff}}}
\end{equation}
cannot be directly computed at $m=0$, where it vanishes on the lattice,
so it must be
obtained as the limit $m \to 0$. To reach the correct continuum value,
this limit has to be taken simultaneously with the $\beta \to \infty$ one,
keeping the product ${\sqrt \beta} m$ fixed.

This can be easily done with this method, which does not require a separate
simulation of the fermionic contribution
for each pair of parameters $(\beta,m)$.

In Figure 2 we report the value of the chiral condensate for three
values of the
ratio $m_c \over e_c$.
For relatively large values of this ratio, scaling sets up already near
$\beta \sim 1$, but even for a very small value (in this lattice) as
${m_c \over e_c} = 0.01$, where finite spacing and
volume effects appear in the small
and large $\beta$ regions, there is a clear scaling window.
We have repeated this procedure for $12$ values of $m_c \over e_c$, and the
values of the chiral condensate in the scaling window
so obtained have been reported in Figure 3.
The behaviour of the condensate is very clear towards the continuum value,
indicated in the figure as a circle.
By fitting the points at small $m_c \over e_c$ with
polynomials
we have always obtained consistent results for the intercept:
\begin{equation}
\langle \bar \psi \psi \rangle = 0.160 \pm 0.002
\end{equation}
in perfect agreement with the theoretical value.

One can exploit the analogy between the fermionic system and a
magnetic one to predict that the chiral condensate should behave as a power
in $m_c \over e_c$ in the scaling region, so allowing an unambiguous
extrapolation at $m=0$.

{}From a formal point of view this result is also important, since it shows
that, even with staggered fermions, where it cannot be proven rigorously,
the usual introduction of the flavour number through powers of the fermionic
determinant is correct; in fact the numerical value for the chiral condensate,
which exactly matches the continuum value, is obtained here
by taking the square root of the determinant in the partition function.

We have also analyzed the zero and two flavours cases \cite{SCHL}. In the zero
flavour limit there is really no scaling region, with the chiral condensate
increasing at large $\beta$, indicating that it diverges as expected
\cite{ZEROF}.
On the contrary in the two flavour case,
the behaviour of the chiral condensate at finite mass indicates a vanishing
value in the chiral limit, again in agreement with expectations \cite{GRADY}.

In conclusion, the results we find agree completely with
the analytical expectations
of the continuum theory. In this respect we believe that the MFA method can
be applied to lattice models where the continuum limit is approached at
infinite inverse coupling, like QCD.

It is particularly interesting, in view of more ambitious applications, the
ease with which constant physics trajectories can be followed in this
approach: in particular, since the mass dependence of the lattice Dirac
operator has become trivial, it is possible to move in the
$\beta, m$ parameter space at negligible computer cost. It is
useful to remember that also the $n_f$ dependence is trivial \cite{QED4}.

This potentiality has been fully exploited in the Schwinger Model, where
Renormalization Group amounts to simple dimensional analysis and Constant
Physics trajectories can be exactly defined through the whole parameter
space; as a consequence our numerical
results for the chiral condensate are quite independent from the
extrapolation to zero fermion mass and (as shown in Figure 3) are by far the
best available in the literature.

We believe that this potentiality can be used in more realistic theories
like QCD.

All the above simulations have been performed on various Transputer
networks at L'Aquila University, Zaragoza University $(RTN)$,
the bulk on the Transputer Networks of the Theory Group of the
Frascati National Laboratories of the INFN.

This work has been partly supported through a CICYT (Spain) - INFN (Italy)
collaboration.
\endpage

\vskip 1 truecm
\leftline{\bf Figure captions}
\vskip 1 truecm
\leftline{Figure 1 \hskip 0.5 truecm Average plaquette $64^2, m=0, n_f=1$.}
\leftline{Figure 2 \hskip 0.5 truecm Chiral condensate, $64^2, {m_c \over
e_c}=0.01, 0.04, 0.08$}
\leftline{Figure 3 \hskip 0.5 truecm Chiral condensate vs ${m_c\over e_c},
n_f=1$, errors are smaller than symbols.}

\end{document}